\documentclass{aiaa-tc}

\graphicspath{	{./figures/} }

\usepackage{cancel}
\usepackage{multirow}
\usepackage{amsmath}
\usepackage{hyperref}
\usepackage{color}

\usepackage{siunitx}
\usepackage{nomencl}
  \makenomenclature
\usepackage{ifthen}

\usepackage{subfloat}
\usepackage[normalem]{ulem}

\usepackage{cleveref}
\usepackage{tikz}

\newcommand{\charlesx}{CharLES$^X$}

\newcommand\Pran{\mbox{\textit{Pr}}}

\usetikzlibrary{shapes}
\usetikzlibrary{plotmarks}

\definecolor{darkgray}{rgb}{0.5, 0.5, 0.5}
\definecolor{lightgray}{rgb}{0.8, 0.8, 0.8}

\newcommand{\legendline}[1][black]{\mbox{(\tikz[baseline=-0.75ex,color=#1]{\draw[very thick,fill=#1]  (0,0) -- (2ex,0);})}}
\newcommand{\legenddashed}[1][black]{\mbox{(\hspace{-0.02cm}\tikz[baseline=-0.75ex, color=#1, dashed, very thick]{\draw  (0,0) -- (3.4ex,0);}\hspace{-0.03cm})}}
\newcommand{\legenddasheddotted}[1][black]{\mbox{(\hspace{-0.02cm}\tikz[baseline=-0.75ex, color=#1, dashed, very thick]{\draw  (0,0) -- (1.4ex,0);}\hspace{-0.03cm}\tikz[baseline=-0.75ex,color=#1]{\draw[very thick,fill=#1]  (0,0) -- (0.45ex,0);}\hspace{-0.00cm})}}
\newcommand{\legenddotted}[1][black]{\mbox{(\hspace{-0.02cm}\tikz[baseline=-0.75ex,color=#1]{\draw[very thick,fill=#1]  (0,0) -- (0.45ex,0);}\hspace{-0.03cm}\tikz[baseline=-0.75ex,color=#1]{\draw[very thick,fill=#1]  (0,0) -- (0.45ex,0);}\hspace{-0.03cm}\tikz[baseline=-0.75ex,color=#1]{\draw[very thick,fill=#1]  (0,0) -- (0.45ex,0);}\hspace{-0.00cm})}}

\newcommand{\legenddot}[1][white]{\mbox{(\tikz[baseline=-0.5ex]{\draw[fill=#1](0.25ex,0.25ex) circle (0.75ex); })}}
\newcommand{\legendsquare}[1][white]{\mbox{(\tikz[baseline=-0.5ex]{\node[rotate=180,fill=blue,color=#1,fill=black,draw=white,stroke=yellow] at (0.0ex,0.25ex){\pgfuseplotmark{square*}};}\hspace{0.cm})}}

\newcommand{\legendtriangles}[1][black]{(\tikz[baseline=-0.4ex]{\hspace{-0.00cm}\node[rotate=180] at (0.0ex,0.25ex) {\pgfuseplotmark{triangle*}};}\hspace{-0.00cm})}

\newcommand{\legendx}[1][black]{($\times$)}

\crefformat{figure}{figure~#2#1#3}
\Crefformat{figure}{Figure~#2#1#3}
\crefmultiformat{figure}{figures~#2#1#3}{ and~#2#1#3}{, #2#1#3}{ and~#2#1#3}
\Crefmultiformat{figure}{Figures~#2#1#3}{ and~#2#1#3}{, #2#1#3}{ and~#2#1#3}
\crefformat{equation}{eq.~(#2#1#3)}
\Crefformat{equation}{Eq.~(#2#1#3)}
\crefformat{section}{\S#2#1#3}
\Crefformat{section}{\S#2#1#3}
\crefmultiformat{section}{\S#2#1#3}{ and~#2#1#3}{, #2#1#3}{ and~#2#1#3}
\Crefmultiformat{section}{\S#2#1#3}{ and~#2#1#3}{, #2#1#3}{ and~#2#1#3}

\title{High-fidelity simulation of an ultrasonic standing-wave thermoacoustic engine with bulk viscosity effects}

\author{
	Jeffrey Lin%
	\thanks{Graduate Student, Department of Electrical Engineering.}\\
	{\normalsize\itshape
		Stanford University, Stanford, CA, 94305, USA}
	\and
	Carlo Scalo%
	\thanks{Professor, School of Mechanical and Aeronautical Engineering. AIAA Member.}\\
	{\normalsize\itshape
		Purdue University, West Lafayette, IN, 47907, USA}
	\and
	Lambertus Hesselink%
	\thanks{Professor, Department of Electrical Engineering. AIAA Member.}\\
	{\normalsize\itshape
		Stanford University, Stanford, CA, 94305, USA}
}

 \AIAApapernumber{YEAR-NUMBER}
 \AIAAconference{55th AIAA Aerospace Sciences Meeting, AIAA SciTech Forum, January 9--13, 2017, Grapevine, TX}
 \AIAAcopyright{\AIAAcopyrightD{2016}}

\begin{document}

\maketitle
\begin{abstract}
We have carried out boundary-layer-resolved, unstructured fully-compressible Navier--Stokes simulations of an ultrasonic standing-wave thermoacoustic engine (TAE) model. 
The model is constructed as a quarter-wavelength engine, approximately 4 mm by 4 mm in size and operating at 25 kHz, and comprises a thermoacoustic stack and a coin-shaped cavity, a design inspired by Flitcroft and Symko (2013).\cite{FlitcroftS_Ultrasonics_2013}
Thermal and viscous boundary layers (order of 10 $\mathrm{\mu}$m) are resolved. 
Vibrational and rotational molecular relaxation are modeled with an effective bulk viscosity coefficient modifying the viscous stress tensor. 
The effective bulk viscosity coefficient is estimated from the difference between theoretical and semi-empirical attenuation curves.
Contributions to the effective bulk viscosity coefficient can be identified as from vibrational and rotational molecular relaxation. 
The inclusion of the coefficient captures acoustic absorption from infrasonic ($\sim$10 Hz) to ultrasonic ($\sim$100 kHz) frequencies. 
The value of bulk viscosity depends on pressure, temperature, and frequency, as well as the relative humidity of the working fluid. 
Simulations of the TAE are carried out to the limit cycle, with growth rates and limit-cycle amplitudes varying non-monotonically with the magnitude of bulk viscosity, reaching a maximum for a relative humidity level of 5\%. 
A corresponding linear model with minor losses was developed; the linear model overpredicts transient growth rate but gives an accurate estimate of limit cycle behavior. 
An improved understanding of thermoacoustic energy conversion in the ultrasonic regime based on a high-fidelity computational framework will help to further improve the power density advantages of small-scale thermoacoustic engines. 
\end{abstract}

\section{Introduction}

\subsection{Background} 
Thermoacoustic engines (TAEs) are devices capable of converting external heat sources into acoustic power, which in turn can be converted to mechanical or electrical power. 
TAEs do not require moving parts and are thermoacoustically unstable past a critical heat input; given this condition, an initial perturbation is sufficient to generate acoustic power amplification. 
The acoustic nature of wave energy propagation in TAEs guarantees close-to-isentropic stages in the thermoacoustic energy conversion process, suggesting the possibility for high efficiency external heat engine designs. 
For example, advanced TAEs have achieved thermal-to-acoustic energy conversion efficiency of 32\%, corresponding to 49\% of Carnot's theoretical limit. \cite{TijaniS_2011}
There are a variety of TAEs in use for energy production and heat pumping, with varying sizes, and heat sources and energy extraction strategies. \cite{Swift_1988}

In a TAE, understanding the energy conversion process from thermal to acoustic is crucial. 
The latter is fluid dynamic in nature and understood and predictable at various levels of fidelity, from quasi one-dimensional linear acoustics \cite{Rott_1980} to fully compressible three-dimensional Navier--Stokes models. \cite{ScaloLH_2015}
However, energy production and dissipation in high-amplitude or high-frequency devices are difficult to model without high-fidelity simulations. 
Similitude offers some answers, but may break down as certain assumptions are violated. \cite{OlsonS_1994}
In the following, we demonstrate a computational modeling strategy, building upon a Navier--Stokes solver, which can simulate high-frequency thermoacoustic amplification with high fidelity.

Modern research has been focused on achieving conversion efficiencies comparable to theoretical expectations. 
Ceperley realized that the thermodynamic cycle induced by engines with traveling wave phasing comprises discrete stages of compression, heating, expansion, and cooling; in standing-wave engines, these stages are partly overlapped and lead to lower energy conversion efficiencies. \cite{Ceperley_1979}
However, Ceperley was unsuccessful in developing a working traveling-wave TAE; the first practical realization is attributed to Yazaki et al. \cite{YazakiIMT_PhysRevLett_1998}
TAEs can therefore largely be classified into standing-wave and traveling-wave configurations, the latter being often more efficient but more complicated to build. 
Hybrid configurations are also possible, with the two concepts combined in a cascaded system. \cite{GardnerS_2003}

Modern thermoacoustic engines are generally large and commonly have operating frequencies between 60 and 400 Hz. 
This design space is constrained by design considerations such as construction limitations and electroacoustic transducer efficiency. 
However, there are significant benefits to engines operating at higher frequencies. 
Power density for standing-wave engines scales favorably with frequency and pressure amplitude, holding constant the operating temperature range.\cite{Swift_2002}. 
Further, acoustic-to-electric conversion via piezoelectric transducers can be more efficient at higher frequencies.\cite{AntonS_SmartMaterStruct_2007, Priya_JElectroceram_2007, ChenXYS_NanoLett_2010}

Acoustic simulations in the ultrasonic regime present some unique challenges. 
Loss mechanisms from thermodynamic non-equilibrium, which can be neglected at lower frequencies, can become dominant\cite{Pierce_2007}, making Stokes's hypothesis invalid as a result. 
Moreover, acoustic streaming may no longer be assumed to be a second-order flow quantity.\cite{MoudjedBHBG_2014}
Oscillating viscous boundary layers may no longer evolve in the continuum regime, and the onset of slip-flow near the wall may be possible. 
As a result, wall-heat transfer may have longer timescales, limiting the intensity of the thermoacoustic response.\cite{FlitcroftS_Ultrasonics_2013}

Previous high-fidelity efforts by Scalo et al.\cite{ScaloLH_2015} demonstrated a full-scale three-dimensional simulation of a large $\sim$60 Hz TAE, revealing the presence of transitional turbulence and providing support for direct low-order modeling of acoustic nonlinearities such as Gedeon streaming. 
More recently, Lin et al.\cite{LinSH_2016} carried out high-fidelity modeling of a $\sim$390 Hz mid-size piezoelectric thermoacoustic energy harvester with particular emphasis on time-domain modeling of electromechanical transmittance functions.
The goal of the current work is to tackle the modeling of a miniature, realistic ultrasonic $\sim$25 kHz engine by accurately capturing ultrasonic attenuation concurrently with thermoacoustic instability, with the support of both high-fidelity and low-fidelity numerical prediction tools. 

\subsection{Research Aims}

In this paper we present a high-fidelity fully compressible Navier--Stokes simulation of a thermoacoustic engine operating in the ultrasonic frequency regime. 
A model for bulk viscosity, incorporating rotational and vibrational relaxation effects, has been developed;
these effects are not insignificant compared with Stokesian viscous and thermal  dissipation\cite{Pierce_1989} and thermoacoustic energy production.
The engine design is based on a standing-wave engine construction first presented by Flitcroft and Symko \cite{FlitcroftS_Ultrasonics_2013} and was chosen due to its simple design. 
To the authors' knowledge, this design is the sole experimental example of an ultrasonic TAE in literature. 
Being able to capture thermoacoustic onset and nonlinear effects are preliminary steps towards the development of computational tools to better predict and optimize energy generation for miniaturized and high-frequency thermoacoustic engines. 

In the following, the adopted theoretical TAE model is first introduced, together with the governing equations and computational setup (\cref{s:model}). 
A technique to capture bulk viscosity and a setup for absorption verification is presented (\cref{s:absorptionsetup}). 
A linear thermoacoustic model predicting the onset and growth of oscillations in the TAE model is presented (\cref{s:linearmodel}). 
Finally, results for the thermoacoustic engine model for both the linear and Navier--Stokes models are shown and discussion follows (\cref{s:discussion}).

\section{Engine Model}
\label{s:model}

\subsection{Engine Model Design and Computational Setup}

\begin{figure}[htb]
	\center
	\includegraphics[width=0.85\textwidth]{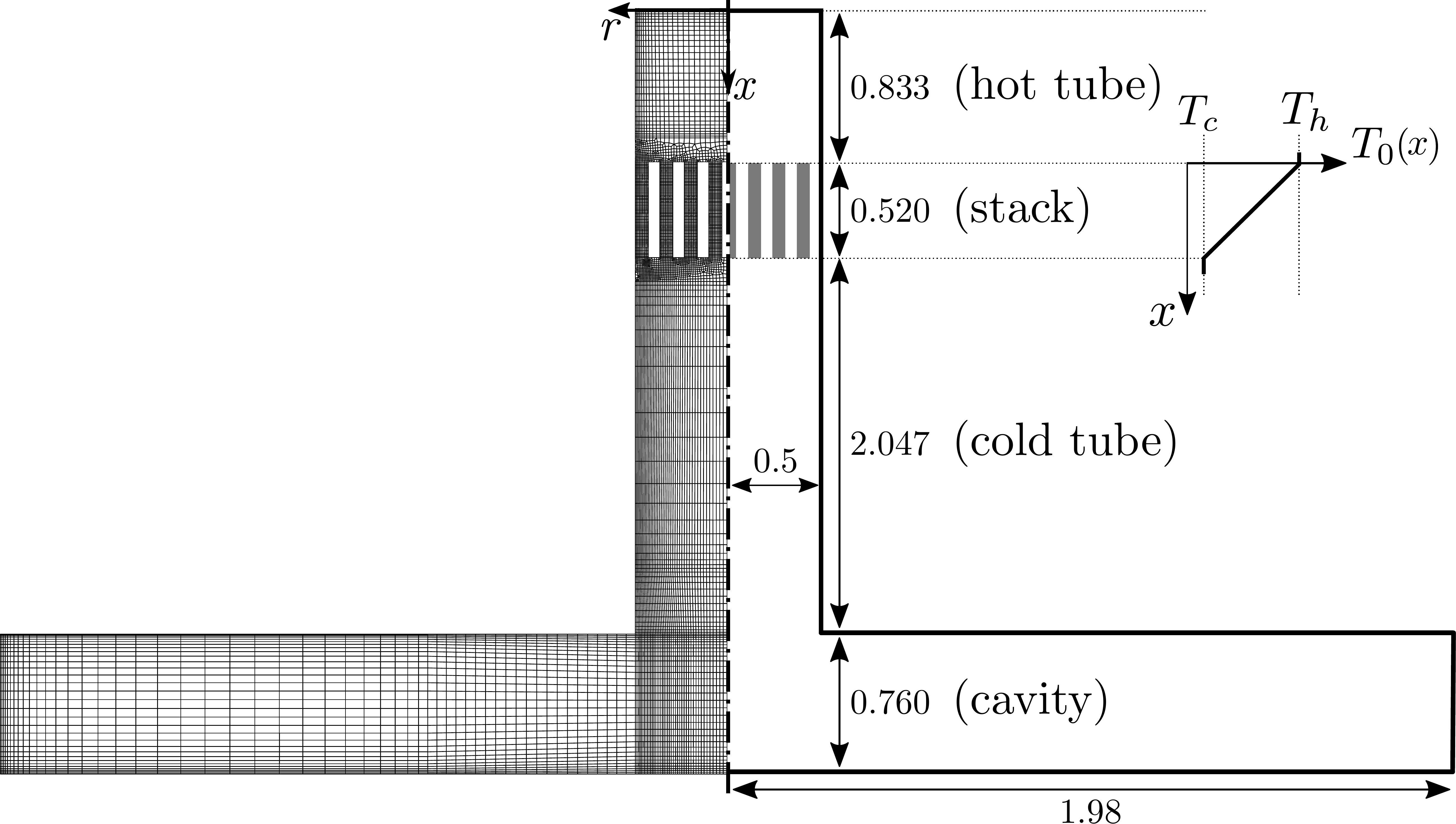}
	\caption{		Computational setup of the axisymmetric standing-wave ultrasonic  TAE model, inspired by Flitcroft and Symko\cite{FlitcroftS_Ultrasonics_2013}  (right half), and sample computational grid (left half). 
		All dimensions are provided in millimeters. 
		The computational mesh is rotationally extruded about the $x$ axis. 
		Geometric properties of the stack are as follows: stack porosity $\phi_s = 0.6$, stack layer count $n_s = 7$, and the stack gap width $h_g = 0.0706$ mm. 
	}
	\label{f:computational_grid}
	\label{f:computational_setup}
\end{figure}

The chosen computational setup (\cref{f:computational_setup}) is a simplified axisymmetric model of the miniature standing-wave thermoacoustic engine first presented by Flitcroft and Symko\cite{FlitcroftS_Ultrasonics_2013}.
Our model is a quarter-wavelength resonator with a thermoacoustic stack in a straight circular tube, closed on one end and connected to a coin-shaped cavity on the other. 
The cavity lowers the critical temperature necessary for onset and also provides for the possibility of pressurizing the engine, and the cavity geometry is estimated from a presentation by Flitcroft and Symko\cite{FlitcroftS_2012}. 
The stack temperature profile varies from $T_c$ of 300 K to $T_h$ of 600 K in our simulations.

The referenced literature reports minimal geometrical information, with only the diameter and length of the tube being explicitly provided. 
As a result, the stack position and the diameter and height/width of the cavity were redesigned such that the engine achieves onset of thermoacoustic instability. 
The stack is constructed as radially-concentric plates, as in Lin et al.,\cite{LinSH_2016} with porosity $\phi_s = 0.6$. 
The number of concentric stack elements ($n_s$) was chosen to be 7, including the centered cylindrical rod, resulting in a stack gap width $h_g$ of 0.0706 mm. 
Linear approximations, using Rott's wave equations, suggest an operating frequency of approximately 21 kHz, while the fully nonlinear Navier--Stokes simulations suggest an operating frequency of approximately 25 kHz. 
Because the defined cavity volume may be different from the experimental setup, the simulation-derived frequency is not expected to match the reported experimental operating frequency of 21 kHz. 

The computational grid, as also shown in \cref{f:computational_grid}, is axisymmetric and designed to resolve thermoviscous boundary layers.
Rotational extrusion of five layers, each of one degree, along the $x$ axis is used to construct the three-dimensional computational grid. 
Adiabatic slipwall conditions are used to impose axial symmetry. 

The high-fidelity model was run both without bulk viscosity (reference) and varying levels of bulk viscosity, as tuned by relative humidity. 
For presented results, several cases were run, differing by gas attenuation magnitude. 
As discussed in detail in \cref{s:relaxationmodifications}, the relative humidity for atmospheric air varies the effective bulk viscosity of the fluid significantly.

\section{Governing Equations and Bulk Viscosity Model}

\subsection{Fully compressible Navier--Stokes equations}

The conservation equations for mass, momentum, and energy, solved in the fully compressible Navier--Stokes simulations of the presented TAE model are, respectively, 
\begin{subequations}
	\label{eq:navierstokes}
	\begin{align}
	\frac{\partial}{\partial t} \left(\rho\right) &+ \frac{\partial}{\partial x_j} \left(\rho u_j \right)  = 0
	\label{subeq:ns1}
	\\
	\frac{\partial}{\partial t} \left(\rho u_i\right) &+ \frac{\partial}{\partial x_j} \left(\rho u_i u_j\right)  =  -\frac{\partial}{\partial x_i} p  +
	\frac{\partial}{\partial x_j} \tau_{ij}
	\label{subeq:ns2}
	\\
	\frac{\partial}{\partial t} \left(\rho \, E\right) &+ \frac{\partial}{\partial x_j} \left[ u_j \left(\rho \, E + p \right) \right] =
	\frac{\partial}{\partial x_j } \left(u_i \tau_{ij} - q_j\right)
	\label{subeq:ns3}
	\end{align}
\end{subequations}
where $x_1$, $x_2$, and $x_3$ (equivalently, $x$, $y$, and $z$) are axial and cross-sectional coordinates, $u_i$ are the velocity components in each of those directions, and $p$, $\rho$, and $E$ are respectively pressure, density, and total energy per unit mass. 
The gas is assumed to be ideal, with equation of state $p= \rho \,R_{gas}\, T$ and a constant ratio of specific heats, $\gamma$. 
The gas constant is fixed and calculated as $R_{gas} = p_{\textrm{ref}} \left(T_{\textrm{ref}}\,\rho_{\textrm{ref}}\right)^{-1}$, 
based on the reference thermodynamic density $\rho_{\textrm{ref}}$, 
pressure $p_{\textrm{ref}}$, 
and temperature $T_{\textrm{ref}}$.
The viscous and conductive heat fluxes are, respectively, 
\begin{subequations}
	\label{eq:heatfluxes}
	\begin{align}
	\tau_{ij} &= 2 \mu \left[S_{ij} + \frac{\lambda}{2 \mu} \frac{\partial u_k}{\partial x_k} \delta_{ij} \right]
		\label{subeq:hf1}
	\\
	q_j &= -\frac{\mu\,C_p}{\Pran} \frac{\partial}{\partial x_j} T
		\label{subeq:hf2}
	\end{align}
\end{subequations}
\nomenclature{$\lambda$}{Second viscosity, Pa$\cdot$s}%
where $S_{ij}$ is the strain-rate tensor, given by $S_{ij}=(1/2) \left(\partial u_j/\partial x_i + \partial u_i /\partial x_j \right)$; $\Pran$ is the Prandtl number; and $\mu$ is the dynamic viscosity, given by $\mu = \mu_{\textrm{ref}}\left(T/T_\textrm{ref}\right)^n$, where $n$ is the viscosity power-law exponent and $\mu_{\textrm{ref}}$ is the reference viscosity. $\lambda$ is the second viscosity defined by 
\begin{align}
\label{e:secondviscosity}
\mu_B \equiv \lambda + \frac{2}{3}\mu \, ,
\end{align}
where $\mu_B$ is the bulk viscosity.
One significant advancement in this work is the adoption of a newly developed bulk viscosity model, accounting for both rotational and vibrational molecular relaxation, as outlined in the following section.
Simulations have been carried out with the following gas properties: $\gamma=1.4$, $\rho_{\textrm{ref}} = 1.2\,\textrm{kg m}^{-3}$, $p_{\textrm{ref}} = 101\,325\,\textrm{Pa}$, $T_{\textrm{ref}}=300\,\textrm{K}$, $\mu_{\textrm{ref}}=1.98\times10^{-5} \, \textrm{kg}\, \textrm{m}^{-1} \textrm{s}^{-1}$,  $\Pran=0.72$, and $n=0.76$, valid for air.\cite{DeYiB_InternationalJournalHeatMassTransfer_1990} 

The governing equations are solved using \charlesx{}, a control-volume-based, finite-volume solver for the fully compressible Navier--Stokes equations on unstructured grids, developed as a joint-effort among researchers at Stanford University. \charlesx{} employs a three-stage, third-order Runge-Kutta time discretization and a grid-adaptive reconstruction strategy, blending a high-order polynomial interpolation with low-order upwind fluxes.\cite{HamMIM_2007} 
The code is parallelized using the Message Passing Interface (MPI) protocol and highly scalable on a large number of processors.\cite{BermejoMorenoBLBNJ_2013}

\subsection{Bulk Viscosity Model}
\label{s:relaxationmodifications}

\subsubsection{Model formulation}

In the following, we outline a novel procedure for estimating the bulk viscosity, $\mu_B$, from the absorption coefficient. 
The latter is the measure of wave attenuation over a given traveled distance and has traditionally been of particular interest for atmospheric acoustics. 
The classical absorption coefficient $\alpha_{cl}$ is 
\begin{align}
\label{e:classicalabsorption}
	\alpha_{cl} & = \frac{\omega^2 }{2\rho_0 a_0^3} \left[ \frac{4}{3} \mu + \mu_{B, rot} + \frac{\left(\gamma-1\right)^2\kappa}{\gamma R  }\right] \, ,
\end{align}
\nomenclature{$\alpha_{cl}$}{Classical absorption coefficient, Np/m}%
\nomenclature{$\omega$}{Angular frequency, rads/s}%
\nomenclature{$\rho_0$}{Base density, kg/m$^3$}%
\nomenclature{$a_0$}{Speed of sound, m/s}%
\nomenclature{$\mu$}{Shear viscosity, Pa$\cdot$s}%
\nomenclature{$\gamma$}{Ratio of specific heats}%
\nomenclature{$\kappa$}{Heat conductivity, W/m-K}%
\nomenclature{$R$}{Specific ideal gas constant, J/kg-K}%
where $\omega$ is angular frequency, and $\kappa$ is the heat conductivity. 
However,  multispecies interactions and rotational and vibrational relaxation result in deviations from classic absorption characteristics at various frequency regimes. \cite{Pierce_1989}
When both bulk viscosity and vibrational relaxation contributions are considered, the absorption coefficient $\alpha_a$ is
\begin{align}
\label{e:absorption}
\alpha_a & = \alpha_{cl} + \sum_{k} \alpha_k \\
\alpha_k  & = \frac{1}{2\pi a_0 /\omega} \left(\alpha_k \lambda  \right)_m \frac{2 \omega \tau_k}{1+\left(\omega \tau_k\right)^2} \\
\left(\alpha_k \lambda\right)_m &= \frac{\pi}{2} \frac{\left(\gamma-1\right)c_{v,k}}{c_p} \\
c_{v,k} & = \frac{n_k}{n} R \left(\frac{T_k^*}{T_k} \right)^2 \exp (-T_k^*/T_k) \, ,
\end{align}
\nomenclature{$\mu_B$}{Bulk viscosity, Pa$\cdot$s}%
\nomenclature{$\alpha_a$}{Absorption coefficient, Np/m}%
\nomenclature[B]{$k$}{Species number}%
where the subscript $k$ indicates the contribution from the $k$-th species, $\tau_k$ is the associated relaxation time given by the semi-empirical relationships for relaxation frequencies, as in \cref{e:relaxationfrequencies}, $\left(\alpha_k \lambda \right)_m$ denotes the maximum absorption per wavelength for the $k$-th species, $n_k/n$ is the mole fraction for the $k$-th species, and $T_k^*$ is the characteristic molecular vibration temperature for the $k$-th species. 
Species in air, for example, are that of nitrogen and gas, with corresponding $n_k/n$ of 0.21 and 0.78, respectively. 

In the present work, the bulk viscosity and absorption contributions from rotational and vibrational relaxation are collapsed into one equation, such that
\begin{align}
\label{e:vibrationrelaxation}
\lambda &=  \left(\mu_{B, rot}+\mu_{B, vib}\right) - \frac{2}{3} \mu  = \mu_{B} - \frac{2}{3} \mu \\
\mu_{B, vib} &= \sum_{k}  \left(\frac{2\rho_0 a_0^3}{\omega^2}\right) \alpha_k \\
&= \sum_{k} \left[\frac{p_0}{2\pi} \left(\gamma-1\right)^2 \left(\frac{n_k}{n}  \left(\frac{T_k^*}{T_k} \right)^2 \exp (-T_k^*/T_k)\right) \right] \frac{ f_k}{f_k^2 + f^2} 
\, ,
\end{align}
\nomenclature{$\mu_{B}$}{Effective bulk viscosity, incorporating vibrational relaxation, Pa$\cdot$s}%
where at atmospheric conditions, $\mu_{B, rot} \approx 0.6\mu$, \cite{Pierce_1989} 
the functional form of $f_k$ is dependent only on temperature, and 
thus the form of effective bulk viscosity $\mu_{B}$ is both frequency and temperature-dependent. 

\begin{figure}[htb]
	\centering
	\includegraphics[width=0.8\textwidth]{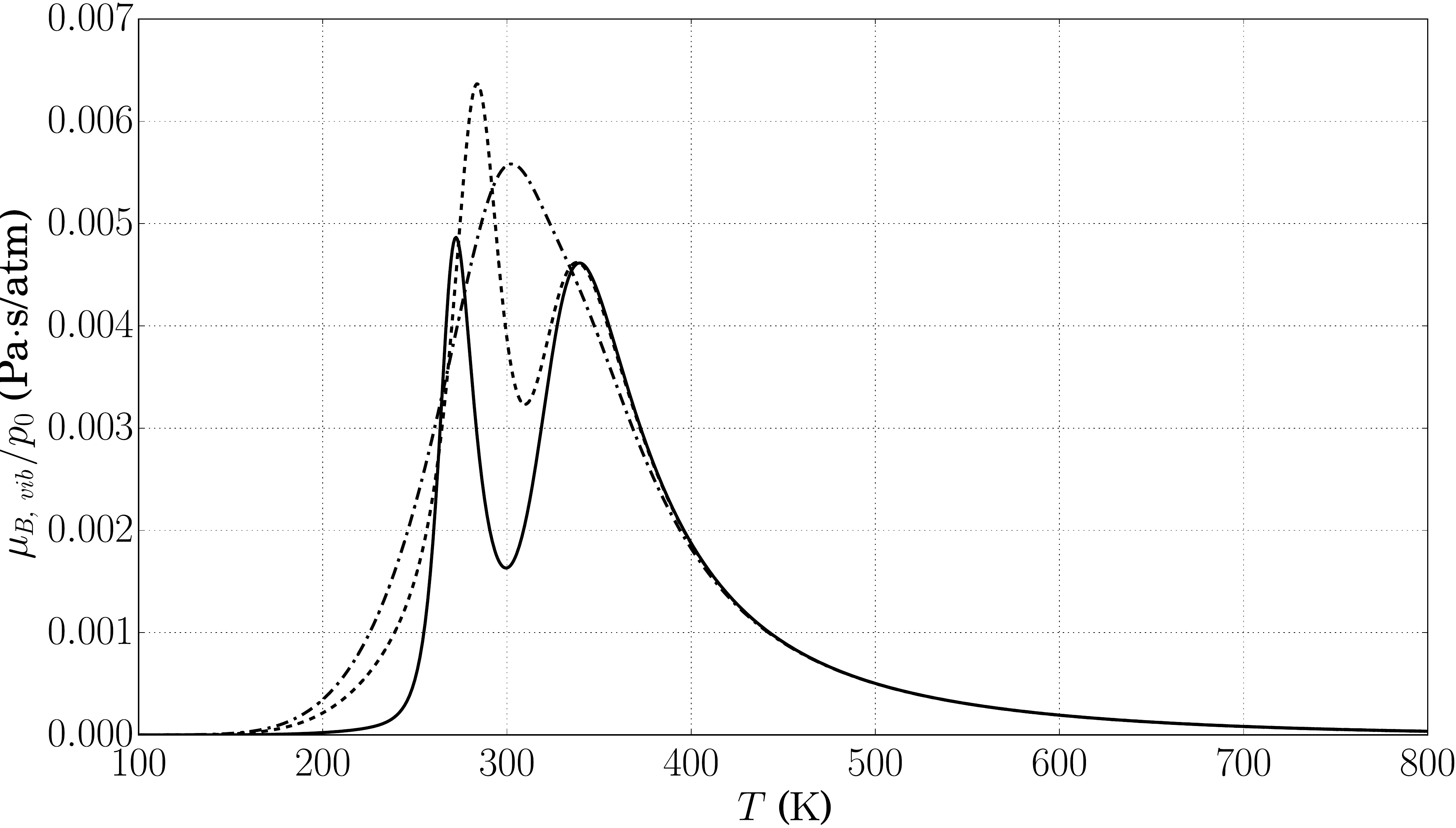}
	\caption{
		Contribution to bulk viscosity from vibrational relaxation, $\mu_{B, vib}$, versus temperature, for pressures $p_0=1$ atm \legendline{}, 10 atm \legenddashed{}, and 100 atm \legenddasheddotted{} and fixed frequency of 1000 Hz. 
	}
	\label{f:mu_B_vs_temp}
\end{figure}

The relaxation frequencies of species in air is given by Bass et al.\cite{BassSZBH_1995}: 
\begin{align}
\label{e:relaxationfrequencies}
\frac{1}{2\pi \tau_{O}} \equiv f_{O} &= \frac{p_0}{p_{s0}} \left(24+4.04\cdot 10^4 h \frac{0.02+h}{0.391+h}\right)\\
\frac{1}{2\pi \tau_{N}}  \equiv f_{N} &= \frac{p_0}{p_{s0}} \left(\frac{T_0}{T}\right)^{1/2} \left(9+280h\cdot \exp \left\{-4.17 \left[ \left(\frac{T_0}{T}\right)^{1/3}-1\right]\right\}\right)\\
\log_{10} \left(p_{\textrm{sat}}/p_{s0}\right) &= -6.8346 \left(T_{01}/T\right)^{1.261}+4.6151 \, ,
\end{align}
\nomenclature{$p_{s0}$}{Reference atmospheric pressure, 101325 Pa}%
\nomenclature{$T_0$}{Reference atmospheric temperature, 293.15 K}%
\nomenclature{$h$}{Absolute humidity, \% $\in\left[0, 100\right]$}%
\nomenclature{$h_r$}{Relative humidity, \% $\in\left[0, 100\right]$}%
\nomenclature{$p_{\textrm{sat}}$}{Saturation vapor pressure, Pa}%
\nomenclature{$T_{01}$}{Triple-point isotherm temperature, 273.16 K}%
where $p_{s0}$ is the reference atmospheric pressure, $p_{\textrm{sat}}$ is the saturation vapor pressure, and $T_{01}$ is the triple-point isotherm pressure. 
The contribution of water vapor to the relaxation frequency of each species in air is determined via an adjustment using the relative humidity $h_r$, 
which defines the absolute humidity $h = h_r \frac{p_\textrm{sat}}{p_0}$. 

\begin{figure}[htb]
	\centering
	\includegraphics[width=0.8\textwidth]{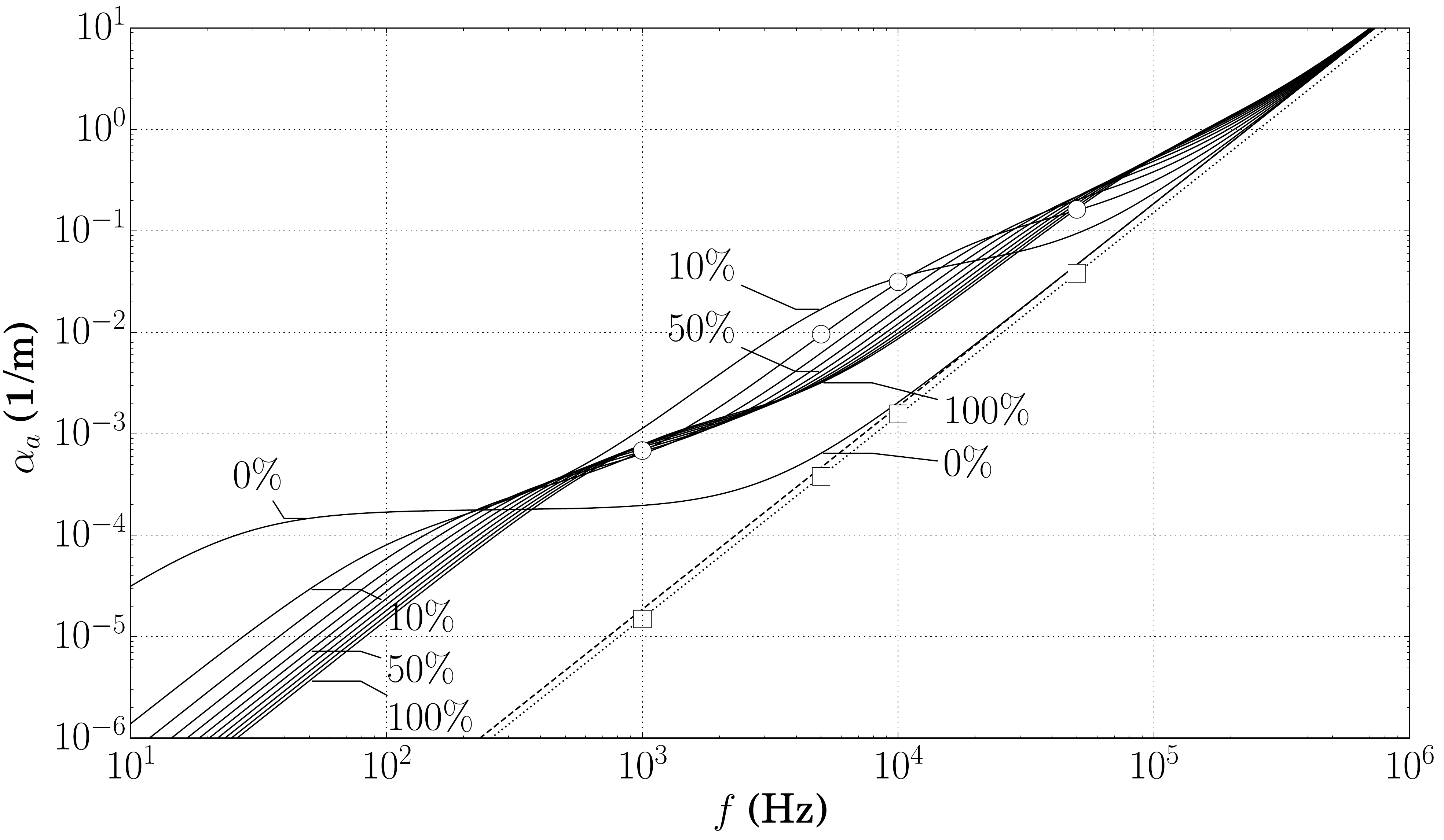}
	\caption{
		Acoustic amplitude attenuation per unit length of propagation in air versus frequency, at 300 K, with semi-empirical expressions \legendline{} (\cref{e:relaxationfrequencies}) and classical expressions with \legenddashed{} and without \legenddotted{} bulk viscosity (\cref{e:classicalabsorption}).
		Relative humidity in percentage is labeled. 
		Computationally-determined absorption for a relative humidity of 20\% are plotted for zero bulk viscosity \legendsquare{} 
		and for calculated effective bulk viscosity \legenddot{}. 
		Absorption coefficient $\alpha_a$ is defined such that $\left|p(x)\right| = \left|p(0)\right| \exp \left(-\alpha_a x \right)$. 
		The absorption relation $-\alpha_a a_0$ can be compared with the thermoacoustic growth rate $\alpha$. 
		Semi-empirical expressions for the relaxation frequencies are provided by Bass et al.\cite{BassSZBH_1995} and equations for the absorption curve are as given in \cref{e:absorption,e:relaxationfrequencies}. 
	}
	\label{f:airattenuation}
\end{figure}

In the vibrational relaxation term, the relaxation times $\tau_k$ are assigned according to relaxation frequencies $f_k = \left(2\pi \tau_k\right)^{-1}$, which have been semi-empirically determined by Bass et al.\cite{BassSZBH_1995}
Sample curves along various relative humidity levels, using the appropriate effective bulk viscosity developed in the preceding equations, are shown in \cref{f:airattenuation}; these curves accurately replicate experimental measurements of absorption in air. 

The form of $\mu_B$ is interpreted in some literature as a frequency-dependent bulk viscosity, as absorption of acoustic power is the primary technique for experimentally measuring the bulk viscosity of a gas. \cite{Emanuel_InternationalJournalEngineeringScience_1998} 
The dependence of $\mu_{B, vib}$ on temperature and pressure is qualitatively depicted in \cref{f:mu_B_vs_temp}. 
The current modeling framework and the ideal gas assumption can break down in dense gas situations; for example, in dense gases, even monatomic gases can exhibit bulk viscosity. \cite{Bhatia_1985}
It is also important to note that an implementation of $\mu_B$ does not allow a solver to fully capture dispersion effects. 
However, because most thermoacoustic engines do not change frequencies significantly in transient and in limit cycle operation, especially if care is taken to minimize thermoacoustic streaming, the predictive ability of this setup is expected to hold in most cases. 

\subsubsection{Time-domain ultrasonic acoustic absorption verification}
\label{s:absorptionsetup}
The modifications made to the governing equations are tested against semi-empirical and analytical expressions for the absorption of sound in air. 

\begin{figure}[htb]
	\centering
	\includegraphics[width=0.85\textwidth]{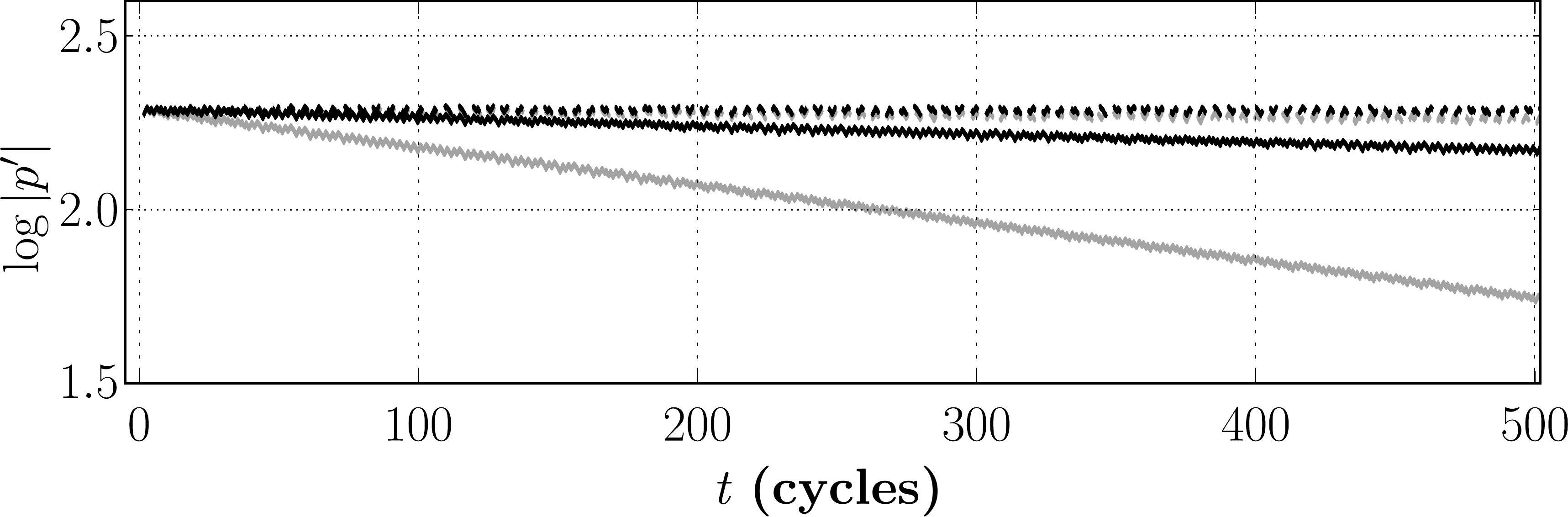}
	\caption{
		Time history of pressure amplitudes of a freely-traveling wave at 300 K, 
		with 1 kHz propagation with zero bulk viscosity \legenddashed[black]{} and with effective bulk viscosity \legendline[black]{}, 
		and for 10 kHz propagation with zero bulk viscosity \legenddashed[gray]{} and with effective bulk viscosity \legendline[gray]{}. 
		Simulation relative humidity in percentage is 20\%. 
	}
	\label{f:attenuation_periods}
\end{figure}

Single-frequency traveling wave one-dimensional simulations with 4096 points per wavelength have been performed in a periodic domain. 
Frequencies in the range $f=10^3-10^5$ Hz have been tested with relative humidity levels of 20\%. 
Due to the monochromatic nature of the wave propagation, the frequency input to the effective bulk viscosity is fixed throughout a single simulation. 
Initial conditions were set as a pure traveling isentropic wave. 
Numerical experiments yielded $\alpha_a$ as annotated in \cref{f:airattenuation}, extracted from time-series decaying pressure amplitudes (\cref{f:attenuation_periods}).

\section{Linear Thermoacoustic Eigenvalue Model}
\label{s:linearmodel}

A system-wide linear model has been developed based on Rott's theory, to support the Navier--Stokes calculations both in the start-up phase and the limit cycle. 
The engine is divided into five Eulerian control volumes, as shown in \cref{f:lsa_grid}: a pre-stack heated duct, the gas-filled volume of the stack, an after-stack duct, an acoustic junction, and the disk-shaped cavity. 
The governing equations have been linearized about the thermodynamic state $\{\rho_0,T_0,P_0\}$. 
The base pressure, $P_0$, is assumed to be uniform, and the mean density and temperature vary with the axial coordinate according to $P_0 = \rho_0(x)\,R_{gas}\,T_0(x)$. 
The base speed of sound is calculated as $a_0=\sqrt{\gamma R_{gas} T_0}$. 
All fluctuating quantities are assumed to be harmonic. 
The $e^{+i\sigma \, t}$ convention is adopted where $\sigma = -i\alpha + \omega$, with $\alpha$ and $\omega$ being the growth rate and angular frequency, respectively.
The thermoacoustic growth rate $\alpha$, measured as $\textrm{s}^{-1}$, can be related to the absorption constant $\alpha_a = \alpha_{cl}+\sum_{k}\alpha_k$ by $-\alpha_a \, a_0$.
\nomenclature{$\alpha$}{Thermoacoustic growth rate, s$^{-1}$}

\subsection{Ducts in the $x$ direction}

In the ducts, a constant axial mean temperature is assumed, yielding the linearized equations
\begin{subequations}
	\begin{align}
	i \sigma \hat{p} &= -\frac{1}{1+\left(\gamma-1\right) f_\kappa} \frac{\rho_0 a_0^2}{A} \frac{d\hat{U}}{d x}  \label{eq:constant_T0_equations_masss_energy} \\
	i \sigma \hat{U} &= -\left(1-f_\nu\right) \frac{A}{\rho_0} \frac{d\hat{p}}{d x} \label{eq:constant_T0_equations_momentum} \, , 
	\end{align} \label{eq:constant_T0_equations}%
\end{subequations}%
which enforce the conservation of mass and energy \eqref{eq:constant_T0_equations_masss_energy} and momentum \eqref{eq:constant_T0_equations_momentum}, respectively. 
The total cross-sectional area is the area available to the gas, $A=A_g$. 
The complex thermoviscous functions $f_\nu$ and $f_\kappa$ in \eqref{eq:constant_T0_equations} are
\begin{equation} \label{eq:complex_viscous_and_thermal_penetration_depth}
f_\nu = \frac{2}{i\,\eta_w} \frac{J_1(i\eta_w)}{J_0(i\eta_w)},	\quad f_\kappa =  \frac{2}{i\,\eta_w\sqrt{\Pran}} \frac{J_1(i\eta_w\sqrt{\Pran})}{J_0(i\eta_w\sqrt{\Pran})}
\end{equation}
where $J_n(\cdot)$ are Bessel functions of the first kind and $\eta$ is the dimensionless complex radial coordinate
\begin{equation}
\eta \equiv \sqrt{\frac{i\omega}{\nu_0}}r =\sqrt{2\,i}\frac{r}{\delta_\nu}
\label{eq:dimensionlessradialcoordinate}
\end{equation}
where $\nu_0=\mu(T_0)/\rho_0$ is the kinematic viscosity based on mean values of density and temperature, and $\eta_w$ in \eqref{eq:complex_viscous_and_thermal_penetration_depth} is the dimensionless coordinate \eqref{eq:dimensionlessradialcoordinate} calculated at the radial location of the isothermal, no-slip wall.
The viscous, $\delta_\nu$, and thermal, $\delta_\kappa$, Stokes thicknesses are
\begin{equation} \label{eq:viscous_thermal_bl}
\delta_\nu = \sqrt{\frac{2\,\nu_0}{\omega }}, \quad \delta_\kappa = \sqrt{\frac{2\,k}{\omega \rho_0 c_p}} \, 
\end{equation}
and are related via the Prandtl number, $\delta_\nu = \sqrt{\Pran} \; \delta_\kappa$. 

\subsection{Thermoacoustic stack in the $x$ direction}

In the thermoacoustic stack, assuming that all annular flow passages share the same pressure field and collapsing via area-weighted averaging leads to a set of approximate linearized equations, 
\begin{subequations}
	\begin{align}
	i \sigma \hat{p}	 &\simeq \sum_{m=1}^{n_s+1} \frac{A_g^{(m)}}{A_g}
	\left[ \frac{\rho_0 a_0^2}{A_g}\frac{1}{1+\left(\gamma-1\right) f_\kappa^{(m)}} \left( \frac{\left(f^{(m)}_\kappa - f^{(m)}_\nu\right)}{\left(1-f^{(m)}_\nu\right)\left(1-\Pran\right)}\frac{1}{T_0} \frac{dT_0}{dx} - \frac{d}{dx} \right)\right]
	\hat{U}
	\label{eq:LST_mass_and_energy_final} \\
	i\sigma \hat{U} &=
	- \sum_{m=1}^{n_s+1}
	\left[ \frac{\left(1-f^{(m)}_\nu\right)A_g^{(m)}}{\rho_0} \frac{d}{dx} \right]
	\hat{p} 	\label{eq:LST_momentum_final}
	\end{align}
	\label{eq:LST_stack_final}
\end{subequations}
where the total cross-sectional area available to the gas, $A_g$, and flow rate, $\hat{U}$, are
\begin{equation}
\quad A_g = \sum_{m=1}^{n_s+1} A_g^{(m)}, \quad A_g^{(m)} = \int_{r^{(m)}_\textrm{bot}}^{r^{(m)}_\textrm{top}} 2\pi\,r\,dr
\end{equation}
\begin{equation}
\hat{U} = \sum_{m=1}^{n_s+1} \hat{U}^{(m)}, \quad \hat{U}^{(m)} = \int_{r^{(m)}_\textrm{bot}}^{r^{(m)}_\textrm{top}} 2\pi\,r\,\hat{u}(r)\,dr
\label{eq:flow_rate_equality}
\end{equation}
and an area-weighted equipartitioning of the flow rates, $\hat{U}^{(m)} =A_g^{(m)}/A_g\; \hat{U}$, has been assumed.
The accompanying thermoviscous functions are 
\begin{subequations}
	\begin{equation}
	\begin{split}
	f^{(m)}_\nu = -\frac{\pi\,\delta_\nu^2}{A_g^{(m)}}
	\Big\{
	\frac{1}{J_0(i\,\eta^{(m)}_\textrm{top})}
	\left[ \eta^{(m)}_\textrm{top} J_1(i\eta^{(m)}_\textrm{top}) - \eta^{(m)}_\textrm{bot} J_1(i\eta^{(m)}_\textrm{bot}) \right] + \\
	\frac{1}{H^{(1)}_0(i\,\eta^{(m)}_\textrm{bot})} \left[ \eta^{(m)}_\textrm{top} H^{(1)}_1(i\eta^{(m)}_\textrm{top}) - \eta^{(m)}_\textrm{bot} H^{(1)}_1(i\eta^{(m)}_\textrm{bot}) \right]
	\Big\}
	\end{split}
	\end{equation}
	\begin{equation}
	\begin{split}
	f^{(m)}_\kappa = -\frac{\pi\,\delta_\kappa^2\,\sqrt{\Pran}}{A_g^{(m)}}
	\Big\{
	\frac{1}{J_0(i\,\eta^{(m)}_\textrm{top}\sqrt{\Pran})}
	\left[ \eta^{(m)}_\textrm{top} J_1(i\eta^{(m)}_\textrm{top}\sqrt{\Pran}) - \eta^{(m)}_\textrm{bot} J_1(i\eta^{(m)}_\textrm{bot}\sqrt{\Pran}) \right] + \\
	\frac{1}{H^{(1)}_0(i\,\eta^{(m)}_\textrm{bot}\sqrt{\Pran})} \left[ \eta^{(m)}_\textrm{top} H^{(1)}_1(i\eta^{(m)}_\textrm{top}\sqrt{\Pran}) - \eta^{(m)}_\textrm{bot} H^{(1)}_1(i\eta^{(m)}_\textrm{bot}\sqrt{\Pran}) \right]
	\Big\} \; , 
	\end{split}
	\end{equation}
\end{subequations}
where $J_n$ and $H_n^{(1)}$ are the Bessel functions of the first kind and Hankel functions of the first kind, respectively. 
A detailed mathematical derivation is found in Lin et al.\cite{LinSH_2016}

\subsection{Cavity in the $r$ direction}

In the radial disk cavity, the governing equations for $\hat{p}$ and $\hat{U}$ vary with perimeter and area as a function of the radius: 
\begin{subequations}
\begin{align}
i\sigma \hat{p} &=-\frac{1}{1+\left(\gamma-1\right) f_\kappa} \frac{\rho_0 a_0^2}{A\left(r\right)} \frac{d\hat{U}}{d r}  \\
i\sigma \hat{U} &= - \left(1-f_\nu\right) \frac{ A \left(r\right)}{\rho_0} \frac{d\hat{p}}{dr}  \, .
\end{align}
	\label{eq:LST_cavity}
\end{subequations}
The thermoviscous functions used in the radial disk are that of parallel plates, 
\begin{align}
f &= \frac{\tanh \left[\left(1+i\right) L/2\delta\right]}{\left(1+i\right)L/2\delta } \, ,
\end{align}
where $L$ is the width of the cavity. 

\subsection{Junction and losses}

In the compliance junction, a relationship on the pressure, input, and output volume flow rates is imposed as 
\begin{align}
i\sigma \hat{p}_J &= \frac{\gamma P_0}{V_J} \left[\hat{U}_x -\hat{U}_r \right] \, , 
\end{align}
where $\hat{U}_x$ and $\hat{U}_r$ are the volume flow rates into and out of the junction, and $V_J$ is the cylindrical volume of the junction. 

Three minor losses---between the duct and stack, stack and duct, and duct and radial cavity---are also accounted for in the linear model in the form of pressure jumps
\begin{align}
\Delta \hat{p}_{ml} &= -\frac{4}{3\pi} \rho \left(\zeta_e + \zeta_c\right) u \hat{u} \, , \label{e:pressuredrop}
\end{align}
where we have adopted the expansion and contraction formulas of Borda-Carnot and Idelchik minor losses: 
\begin{align}
\zeta_e &= \left(1-\frac{A_0}{A_1}\right)^2\\
\zeta_c &= 0.5 \left(1-\frac{A_0}{A_1}\right)^{0.75},
\end{align}
where $A_0$ and $A_1$ are the smaller and larger areas, respectively. 
The pressure drop \cref{e:pressuredrop} is linearized about a given acoustic velocity amplitude, $U$, which is updated in time based on growth rate information, $U=U_0 \exp \left(\alpha t\right)$, where $U_0$ is an initial velocity amplitude. 

Further implementation details can be found in Lin et al.\cite{LinSH_2016}
Linear stability analysis, in the present model, does not incorporate bulk viscosity effects. 

\begin{figure}[htb]
	\begin{center}
\includegraphics[width=0.60\textwidth,trim=2cm 0 0 0]{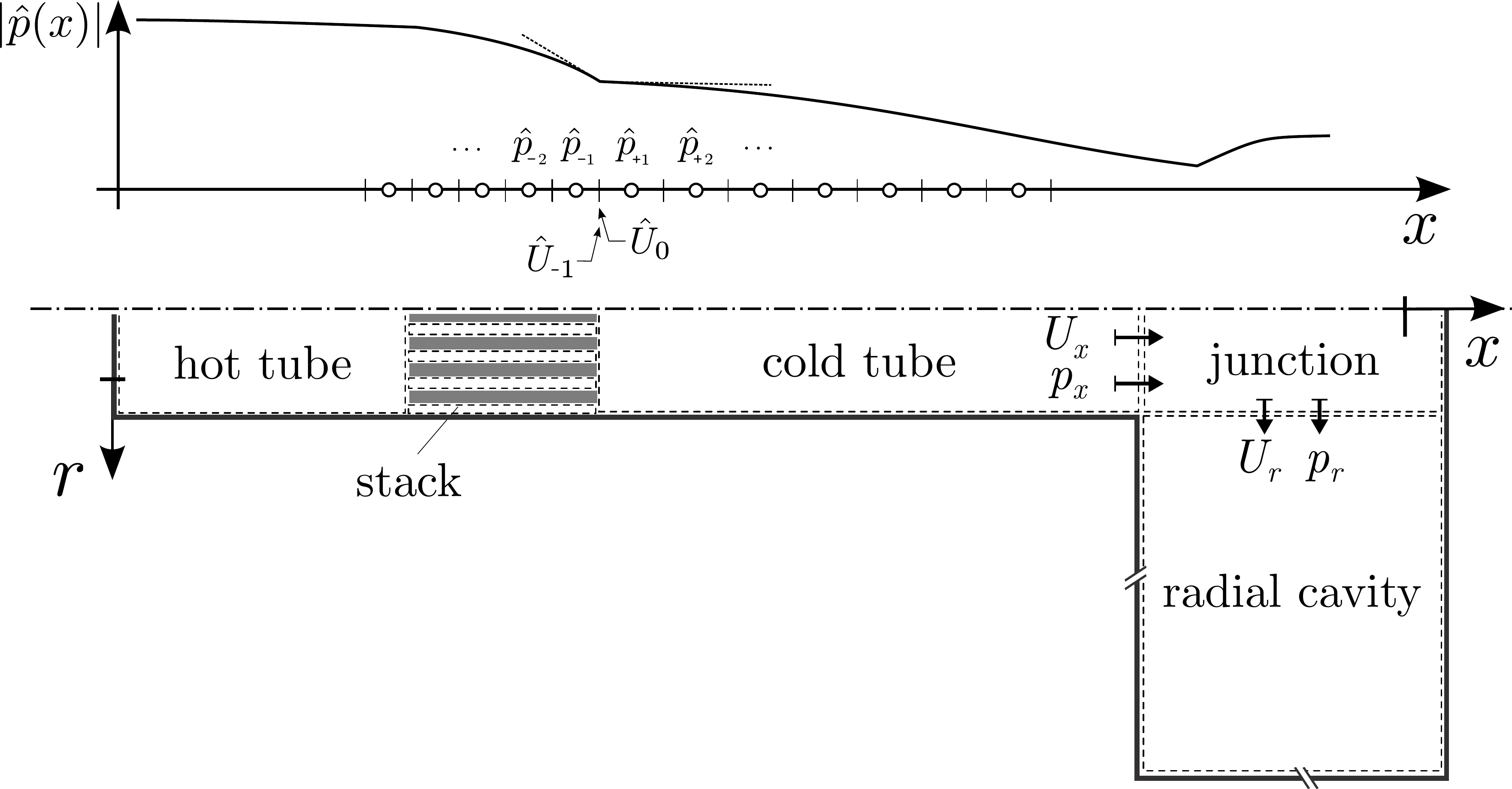}
\includegraphics[width=0.38\textwidth,trim=0cm 0 5cm  8cm]{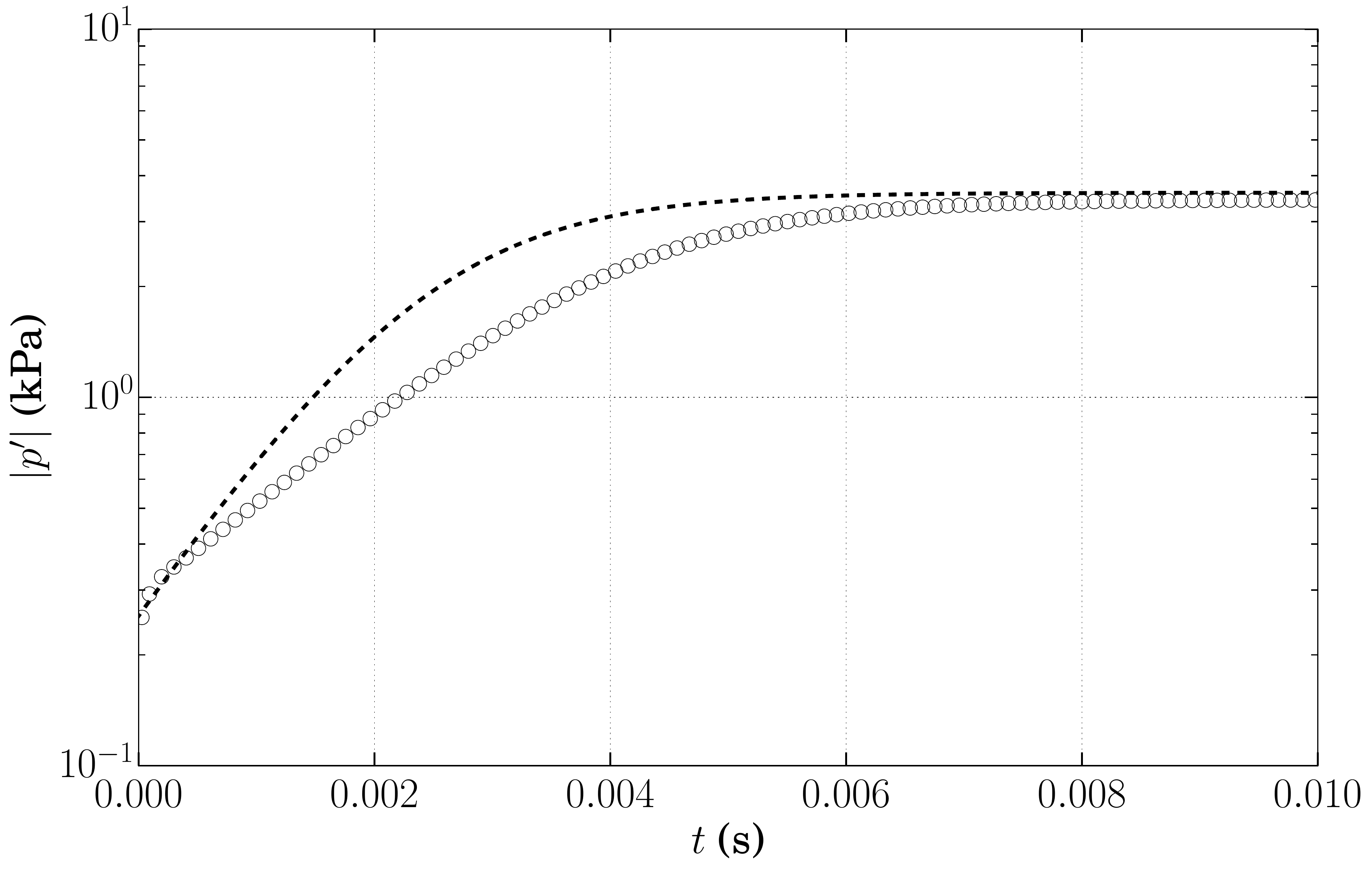}
\end{center}
	\caption{Lumped parameter model used for the linear stability analysis (left). 
		Transient pressure amplitudes of the modeled thermoacoustic engine from Navier--Stokes simulations  \legenddotted{} and linear model \legenddashed{} (right). 
	}
	\label{f:lsa_grid}
	\label{f:transient_growth}
\end{figure}

\section{Results}

\label{s:results}

The thermoacoustic engine model (see \cref{f:computational_setup}) without bulk viscosity effects (the reference case) and with various levels of bulk viscosity, as tuned by working fluid humidity, were simulated through transient growth and thermoacoustic amplification into the first limit cycle. 
Since the effective bulk viscosity implies a direct effect on the pressure gradient and has a non-trivial relationship with pressure, it is also expected that Navier--Stokes computational results will be highly dependent on humidity levels. 

We first present a pressure amplitude timeseries of the results without a bulk viscosity model, as shown in \cref{f:transient_growth}. 
We also have overlaid transient results from the linear model, which reaches limit cycle due to nonlinearities from minor losses. 
The time series for the linear model is integrated along its evolution over $\exp \left(\left(\alpha+i\omega\right)t\right)$. 
While the limit cycle is well-predicted, as minor losses increase with larger flow rate amplitude, the transient growth rate is overpredicted by the linear model. 

\begin{figure}
	\centering
		\begin{minipage}{0.3\textwidth}
			\includegraphics[width=\linewidth]{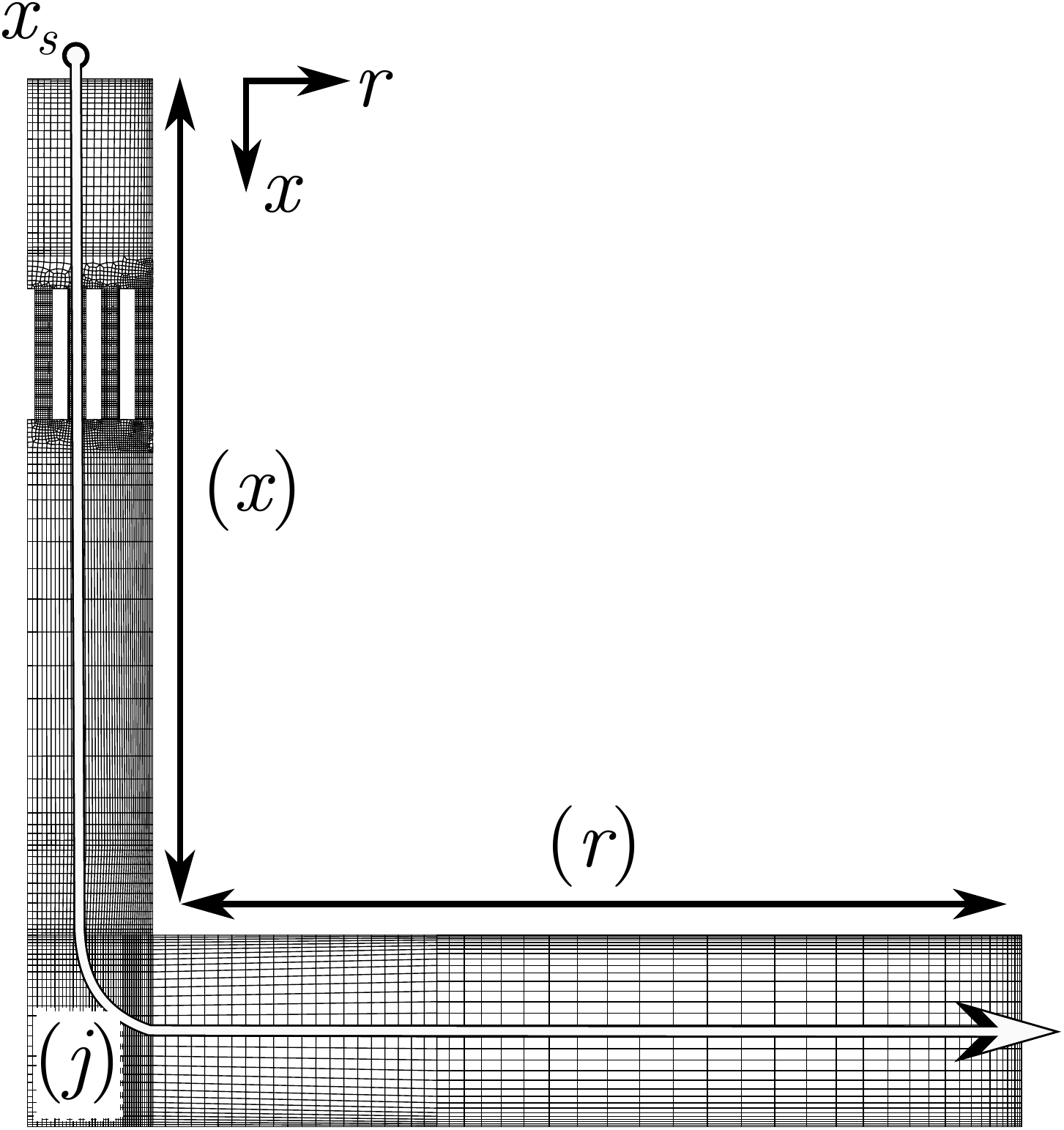}
		\end{minipage}%
		\begin{minipage}{0.65\textwidth}
			\begin{center}
				\includegraphics[width=\linewidth]{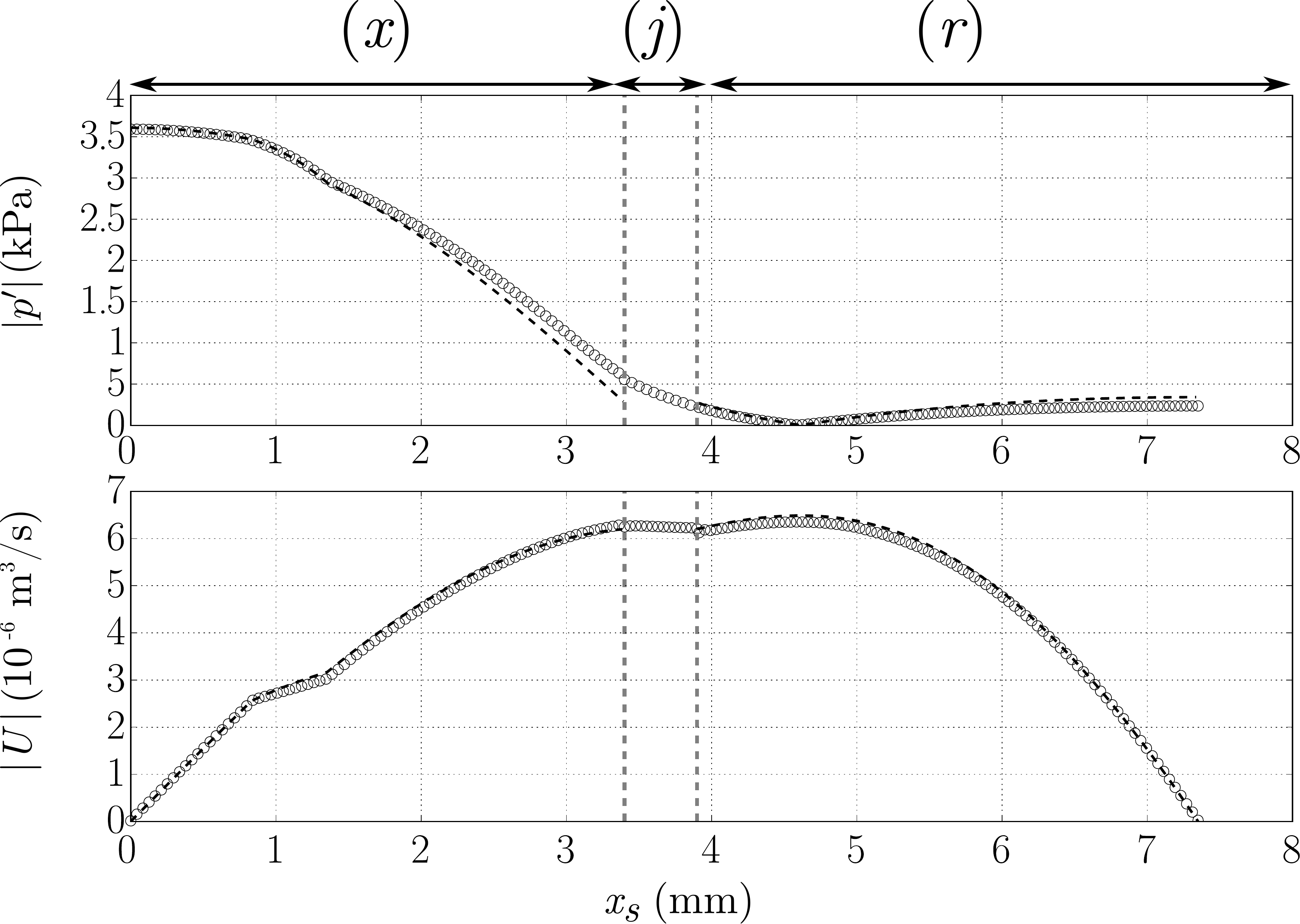}
			\end{center}
		\end{minipage}
	\caption{ 
		Pressure and flow rate fluctuation amplitudes along the curvilinear abscissa $x_s$, as defined by an $x$ segment through the tubes and stack, an abscissa $j$ segment through the junction, and an $r$ segment through the radial cavity. Simulation data  \legenddot{} vs Rott's theory \legenddashed{} along defined axis. }
	\label{f:rotts_profile}
\end{figure}

Acoustic amplitudes along the engine, as defined through the tubes, stack, and radial cavity in \cref{f:rotts_profile}, were also computed for the reference model.
These are shown in \cref{f:rotts_profile} and qualitatively compare well. 

\begin{figure}[htb]
	\centering
	\includegraphics[width=0.8\textwidth]{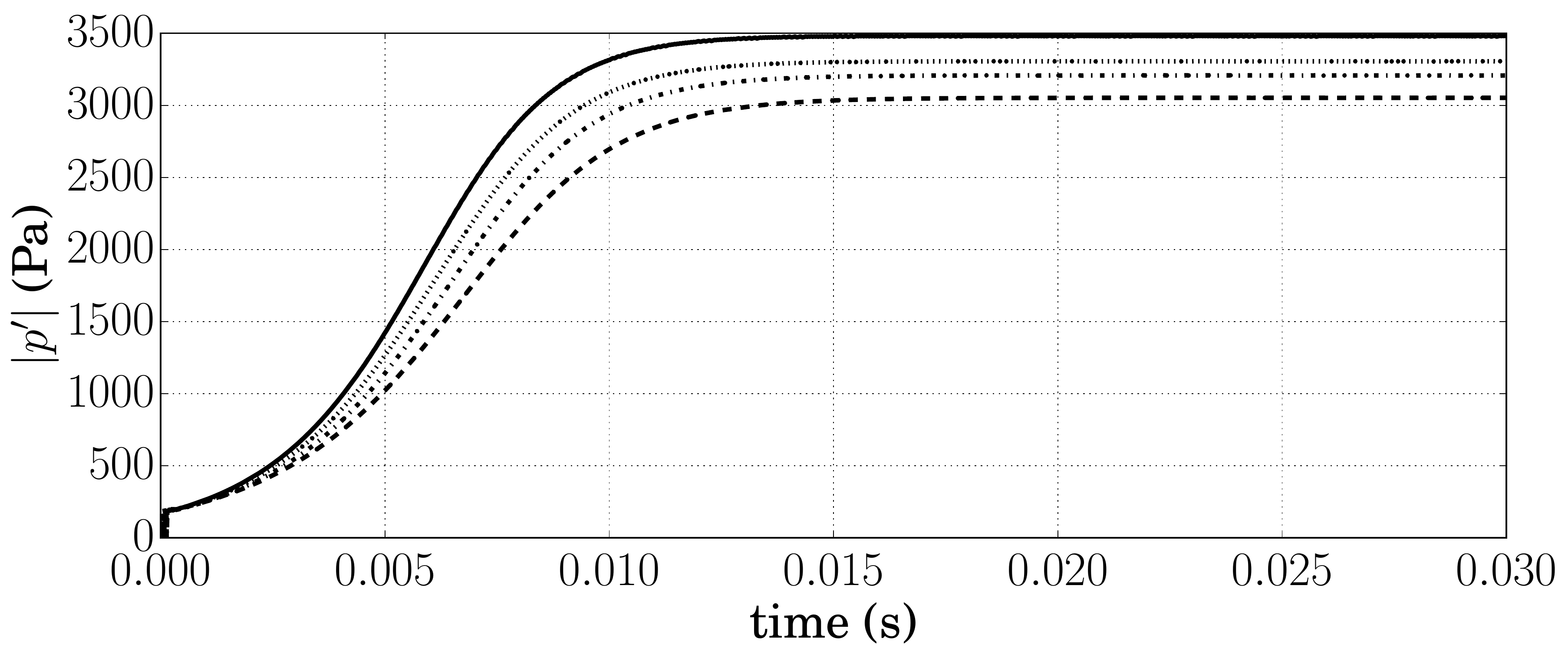}
	\caption{
		Time-series of pressure amplitudes predicted by Navier--Stokes calculations: no bulk viscosity \legendline{} and with bulk viscosity,  corresponding to relative humidity levels of 1\% \legenddashed{}, 5\% \legenddotted{}, and 20\% \legenddasheddotted{}. 
	}
	\label{f:transient_comparison}
\end{figure}

Three additional cases were run, corresponding to a working fluid of atmospheric air with 1\%, 5\%, and 20\% relative humidity. 
The transients for each of these are shown in \cref{f:transient_comparison}. 

Notably, the reference case has the greatest growth rate and limit cycle amplitude, approximately 3503 Pa. 
This is to be expected, as the attenuation from bulk viscosity is not accounted for in the reference case. 
The limit cycle amplitude of the 5\% relative humidity case is calculated to be approximately 3304 Pa, which is significantly higher than that of the 20\% humidity case (3208 Pa) and that of the 1\% humidity case (3054 Pa). 

At the limit cycle, the Navier--Stokes model also reveals the presence of imperfect Helmholtz resonator behavior. 
Compressibility within the system is not neglected, and acoustic energy fluctuates through not only the quarter-wavelength engine sections but also through the radial cavity.

Streaming patterns appear similar to other standing-wave engines thus far studied,\cite{LinSH_2016} and thermoacoustic heat transport away from the stack qualitatively is lower than expected (not shown).

\section{Discussion}
\label{s:discussion}

\subsection{Effective acoustic pressure}

Bulk viscosity has the largest effect near vibrational energy ``peaks,'' where the resonance frequency approaches the natural frequency of a vibrational mode.
In order to evaluate the effect of bulk viscosity under different conditions, we consider its effect on the acoustic effective pressure.
The physical manifestation of the bulk viscosity within the momentum and energy equations, \cref{eq:navierstokes}, is as an adjustment to the thermodynamic pressure, i.e.: 
\begin{align}
p_{\textrm{eff}} &= p - \mu_{B} \nabla \cdot \vec{u}' 
\end{align}
Per linear acoustics and assuming simple oscillations for which pressure and velocity are in standing-wave or traveling-wave phasing, 
\begin{align}
\frac{1}{\rho_0 a_0^2} \frac{\partial p'}{\partial t} + \nabla \cdot \vec{u'} &= 0 \\
\frac{\omega}{\rho_0 a_0^2} \left| p' \right| + \left|\nabla \cdot \vec{u'} \right| &= 0  \, ,
\end{align}
hence resulting in 
\begin{align}
\mu_B^* = \frac{\mu_{B} \omega}{\gamma p_0}  &\approx \frac{-\mu_{B} \left| \nabla \cdot \vec{u}' \right|}{\left| p' \right|} \approx \frac{\left|p'_{\textrm{eff}}\right| - \left|p'\right|}{\left| p' \right|} \, .
\end{align}
This suggests that the dimensionless group $\frac{\mu_{B} \omega}{\gamma p_0}$ is a measure of the relative importance of bulk viscosity effects on the effective acoustic pressure. 

\begin{figure}[htb]
	\begin{center}
		$(a)_{\includegraphics[width=0.46\textwidth]{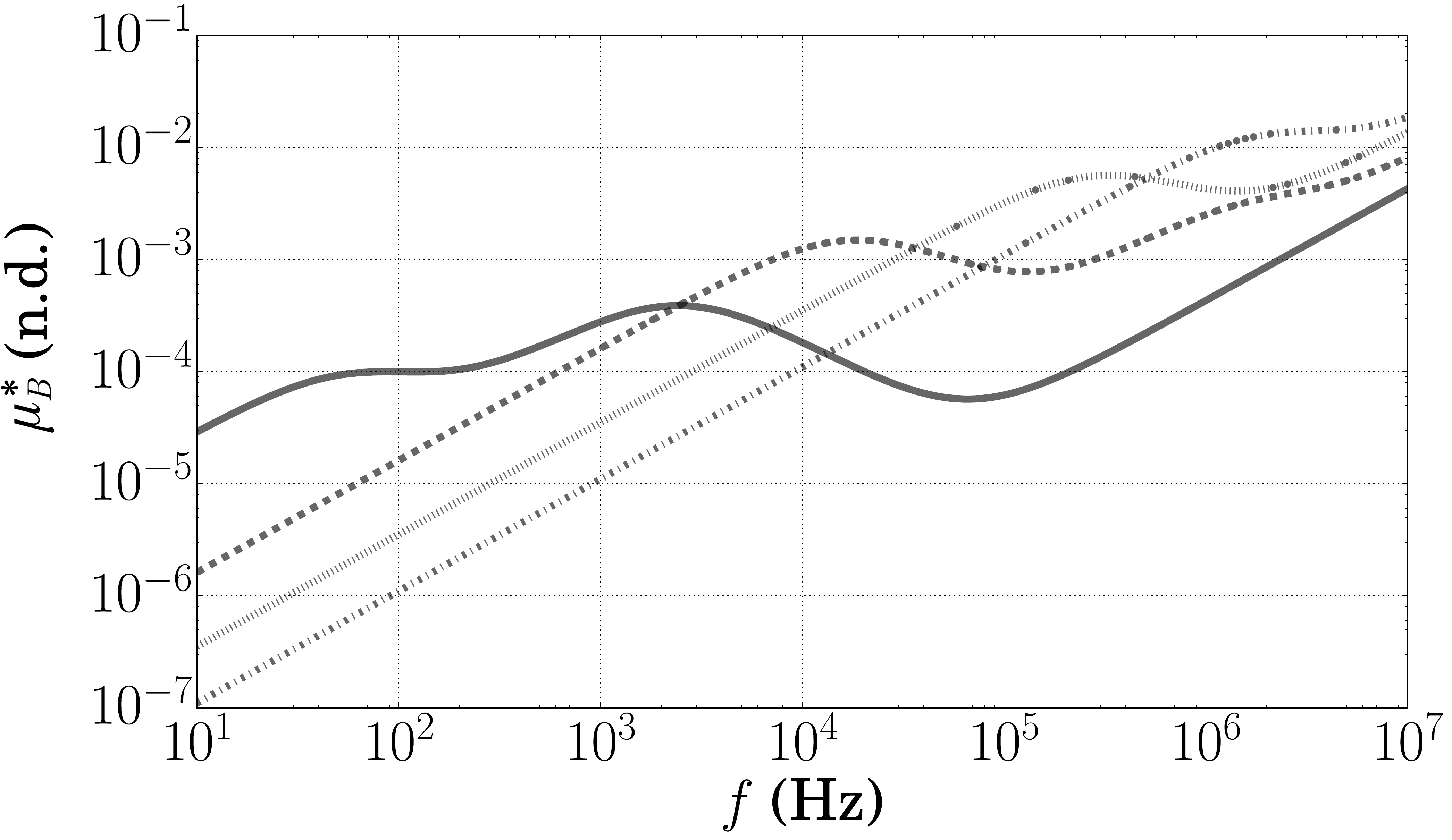}}$ 
		$(b)_{\includegraphics[width=0.46\textwidth]{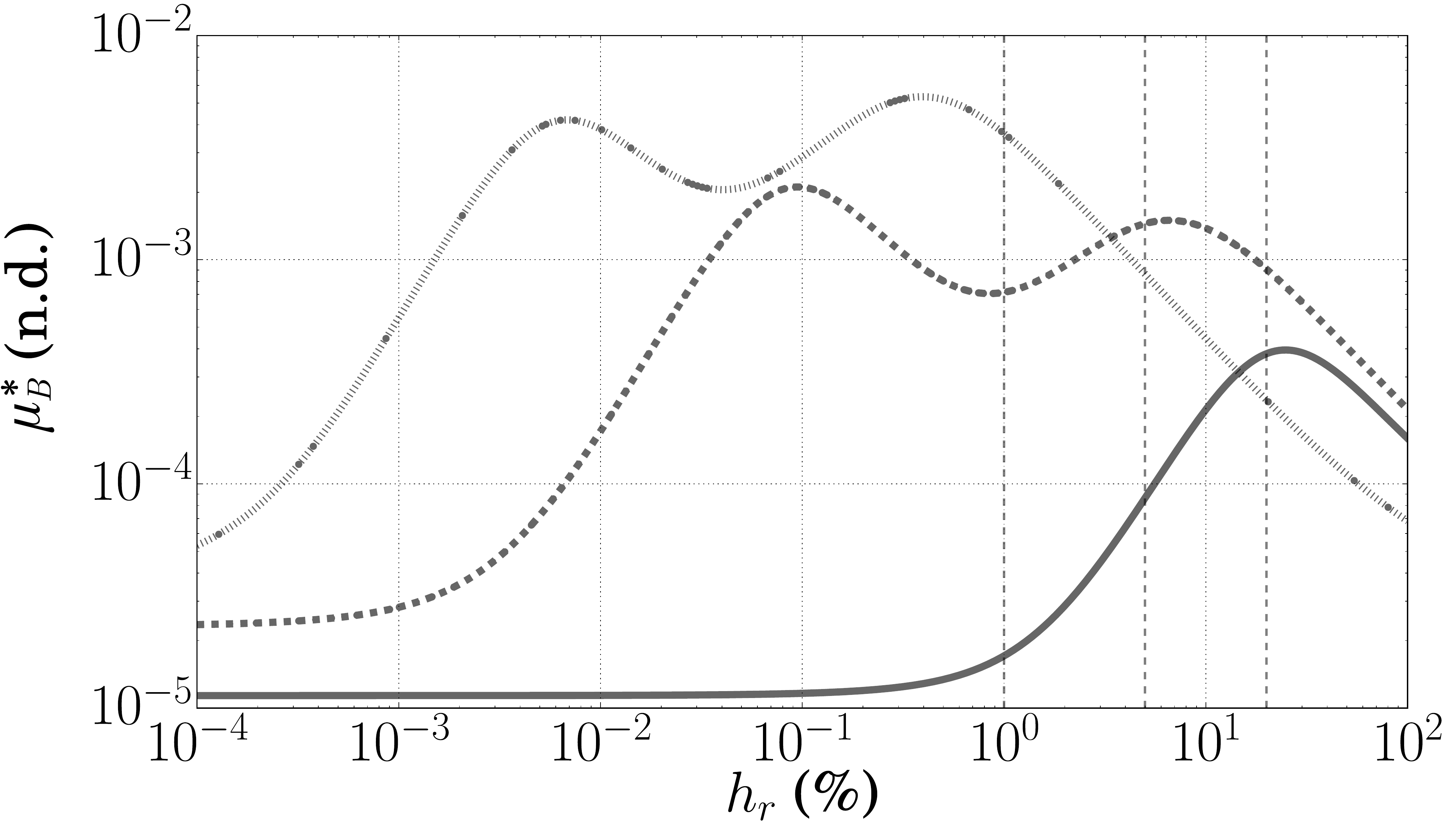}}$ 
	\end{center}
	\caption{
		Dimensionless effective bulk viscosity $\mu_B^*$ versus frequency at $T=$ 300 K \legendline{}, 450 K \legenddashed{}, 600 K \legenddotted{}, and 750 K \legenddasheddotted{}. 
		Results shown for $h_r=5\%$ and atmospheric pressure (a). 
		Dimensionless effective bulk viscosity $\mu_B^*$ versus relative humidity, $h_r$, at $T=$ 300 K \legendline{}, 450 K \legenddashed{}, and 600 K \legenddotted{}.
		Chosen humidity levels of 1\%, 5\%, and 20\% are highlighted with vertical dotted lines (b).  
	}
	\label{f:bulk_nd_with_temp}
\end{figure}

Traditional attenuation curves, measured relative to attenuation per meter, tend to belie the effect of bulk viscosity at high frequencies. 
When measured relative to acoustic wavelength, the bulk viscosity contribution to attenuation is shown to be as large as 1\% of pressure amplitude and has a magnitude peak varying with gas temperature, pressure, and humidity, as seen in \cref{f:bulk_nd_with_temp}. 

\subsection{Contribution of bulk viscosity}

In the previous section, it was shown that the growth rate and limit cycle amplitude vary non-monotonically with the relative working fluid humidity. 
\Cref{f:bulk_nd_with_temp}b suggests one possible explanation of the phenomenon. 
The bulk viscosity contribution to attenuation at the base temperature of 300 K increases monotonically with humidity (for $h_r < 20\%$); as a result, the 20\% relative humidity case has the highest attenuation in the regions of the engine which are at ambient temperature. 
The bulk viscosity contribution to attenuation at the peak temperature of 600 K decreases with humidity; as a result, the 1\% relative humidity case has the highest attenuation in the heated portion of the engine. 
For the optimal case, with 5\% relative humidity, attenuation in the heated and ambient temperature portions of the engine are reduced. 

\begin{figure}[htb]
	\center
	\includegraphics[width=0.8\textwidth]{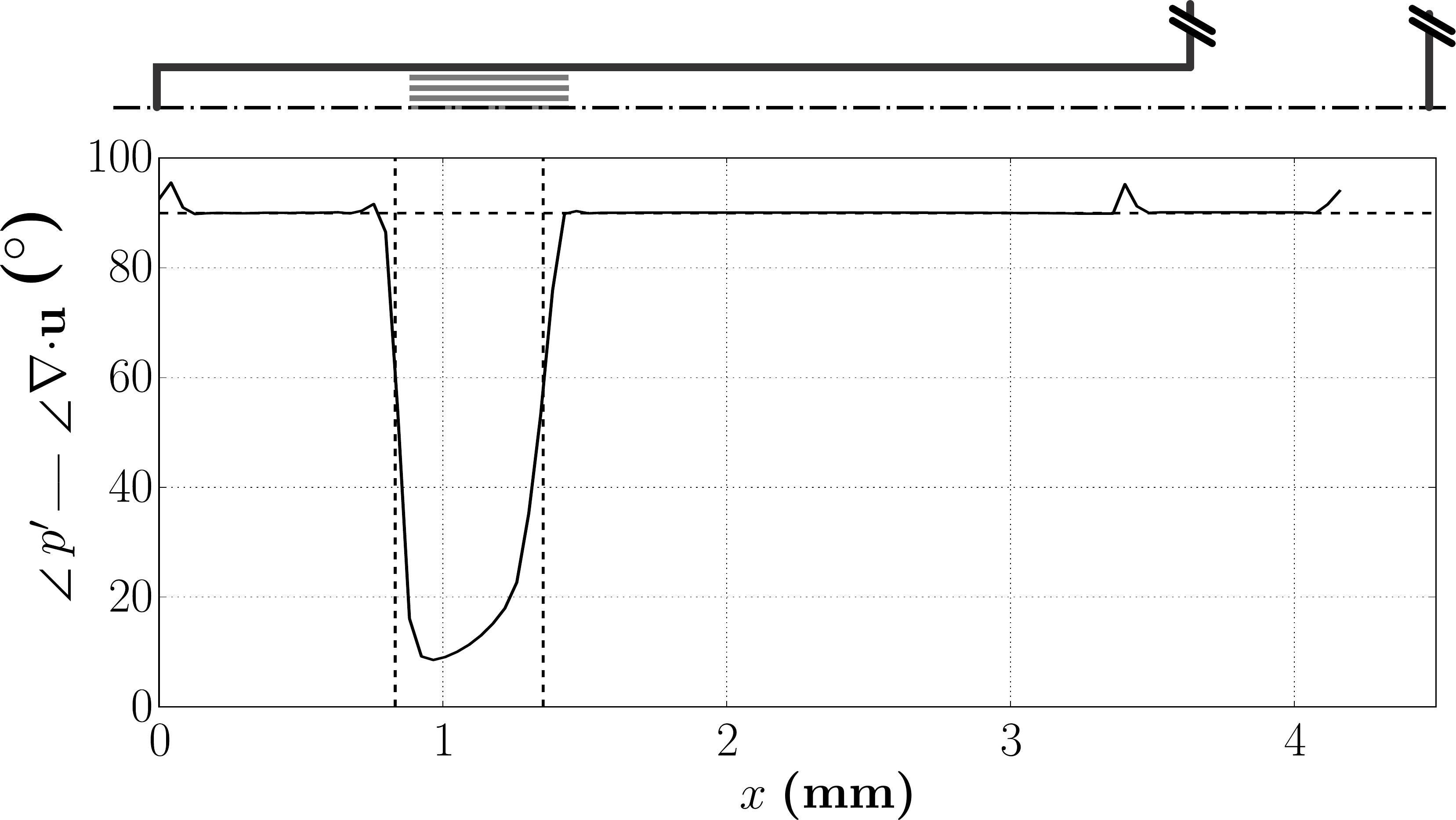}
	\caption{
		Phase difference between pressure fluctuation and the velocity field divergence throughout the engine $x$ axis. 
	}
	\label{f:phasing}
\end{figure}

The phase difference between pressure fluctuations and the velocity field divergence (dilatation) term contributes directly to both acoustic wave attenuation and thermoacoustic energy production. 
For both purely standing- and traveling-waves, pressure fluctuation and dilatation are 90 degrees out of phase. 
In the case of the presented standing-wave engine, it is found that such a phase difference holds everywhere but in the thermoacoustic stack, where it approaches zero (\cref{f:phasing}). 
This is consistent with interpretations of standing-wave thermoacoustic energy production, in which, due to the imposed temperature gradient, a fluid parcel oscillates between high pressure and expansion and low pressure and contraction. 

It is yet unclear whether bulk viscosity directly affects thermoacoustic energy conversion. 
However, these results suggest that there is an opportunity for miniature thermoacoustic engines to be designed with bulk viscosity effects in mind. 
Tracing the Lagrangian fluid parcel within the stack would suggest that bulk viscosity attenuation is minimal within the stack channel and has a much larger effect in the rest of the engine. 
The limit cycle pressure amplitude differs by over 7\% between cases which incorporate bulk viscosity, suggesting a non-negligible effect on the acoustic power output. 

\subsection{Summary and Future Work}

Several results are currently presented. 
A Navier--Stokes code has been shown to accurately capture acoustic absorption under a range of frequencies relevant to both acoustic and thermoacoustic applications. 
A method for evaluating the attenuation strength of bulk viscosity in different conditions is presented. 
The use of bulk viscosity adjustments via temperature, pressure, and humidity provides a computational baseline. 
Further investigation will reveal how relaxation at high-frequency affects thermoacoustic onset, relative to the baseline. 
Results for thermoacoustic amplification have been presented, using a modified version of the Flitcroft and Symko engine. 
As suggested in the paper, limit cycle results also differ significantly between numerics which account for bulk viscosity and for solvers which do not account for bulk viscosity. 

Continuing work include the computation and analysis of additional cases, a generalized predictive model for the effect of $\mu_B$ on thermoacoustic onset and limit cycle behavior, and visual intuition for how an effective bulk viscosity can affect thermoacoustic behavior. 
These results are expected to conclude in techniques for optimization of high-frequency engines.

\bibliography{bibtex_database}
\bibliographystyle{aiaa}

\end{document}